\documentclass{article}
\usepackage{amsmath,amssymb,amsfonts}

\def\op#1{\mathop{{\it\fam0} #1}\limits}

\newcommand{\di}{\mathrm{dim}}

\newcommand{\Ker}{\mathrm{Ker}\,}

\newcommand{\beq}{\begin{equation}}
\newcommand{\eeq}{\end{equation}}
\newcommand{\ben}{\begin{eqnarray}}
\newcommand{\een}{\end{eqnarray}}
\newcommand{\be}{\begin{eqnarray*}}
\newcommand{\ee}{\end{eqnarray*}}
\newcommand{\bea}{\begin{eqalph}}
\newcommand{\eea}{\end{eqalph}}

\newcommand{\gE}{{\mathfrak E}}

\newcommand{\cL}{{\mathcal L}}

\newcommand{\cE}{{\mathcal E}}
\newcommand{\cH}{{\mathcal H}}

\newcommand{\dl}{\delta}
\newcommand{\la}{\lambda}
\newcommand{\La}{\Lambda}

\newcommand{\om}{\omega}
\newcommand{\Om}{\Omega}
\newcommand{\m}{\mu}

\newcommand{\ve}{\varepsilon}

\newcommand{\w}{\wedge}

\newcommand{\wt}{\widetilde}
\newcommand{\wh}{\widehat}
\newcommand{\ol}{\overline}

\newcommand{\dr}{\partial}

\newcommand{\ar}{\op\longrightarrow}

\newcommand{\ot}{\otimes}

\newenvironment{eqalph}{\stepcounter{equation}
\setcounter{equationa}{\value{equation}} \setcounter{equation}{0}

\begin{eqnarray}}{\end{eqnarray}\setcounter{equation}{\value{equationa}}}

\newcommand{\mar}[1]{}

\begin{document}

\hbox{}

\begin{center}

{\Large\bf

Deviation differential equations. Jacobi fields}

\bigskip

G. SARDANASHVILY

\medskip

Department of Theoretical Physics, Moscow State University, 117234
Moscow, Russia

\bigskip

\end{center}

\begin{abstract}
Given a differential equation on a smooth fibre bundle $Y\to X$,
we consider its canonical vertical extension to that, called the
deviation equation, on the vertical tangent bundle $VY$ of $Y$.
Its solutions are Jacobi fields treated in a very general setting.
In particular, the deviation of Euler--Lagrange equations of a
Lagrangian $L$ on a fibre bundle $Y$ are the Euler--Lagrange
equations of the canonical vertical extension of $L$ onto $VY$.
Similarly, covariant Hamilton equations of a Hamiltonian form $H$
are the Hamilton equations of the vertical extension $VH$ of $H$
onto $VY$.
\end{abstract}

\section{Introduction}

By a Jacobi field usually is meant a vector field along a geodesic
in a pseudo-Riemannian manifold which obeys the geodesic deviation
equation.

In a general setting, we consider the deviation equation
(\ref{140}) -- (\ref{141}) and the Jacobi fields of an arbitrary
differential equation (\ref{5.88}) on a smooth fibre bundle $Y\to
X$. In particular, we are concerned with the deviation of the
Euler--Lagrange equations (\ref{151}) and the covariant Hamilton
equations (\ref{b4100}) on a fibre bundle $Y\to X$.

Let $Y\to X$ be a smooth fibre bundle and $J^rY$ the $r$-order jet
manifold of its sections \cite{book09,sard13}. An $r$-order
differential equation on $Y$ conventionally is defined as a closed
subbundle of the jet bundle $J^rY\to X$ \cite{bry,book,kras}. We
restrict our consideration to differential equations which come
from differential operators on $Y$.

Let $E\to X$ be a vector bundle. A $E$-valued $r$-order
differential operator on $Y$ is a bundle morphism
\mar{132}\beq
\Delta: J^rY\ar_X E \label{132}
\eeq
over $X$. Given a global zero section $\wh 0$ of $E\to X$, we
treat its inverse image
\mar{131}\beq
\gE=\Ker \Delta=\Delta^{-1}(\wh 0)  \label{131}
\eeq
as a differential equation on $Y$, though it need not be a closed
subbundle of $J^rY$.

Let $(x^\la,y^i)$ be bundle coordinates on $Y$,
$(x^\la,y^i,y^i_\La)$ the adapted coordinates on $J^rY$, and
$(x^\la,z^A)$ bundle coordinates on $E$. Then the differential
equation $\gE$ (\ref{131}) is locally given by equalities
\mar{5.88}\beq
\cE^A(x^\la, y^i_\La) =0. \label{5.88}
\eeq

Let $VJ^rY$ and $VE$ be the vertical tangent bundles of fibre
bundle $J^rY\to X$ and $E\to X$, respectively. There is the
canonical vertical prolongation
\mar{135}\beq
V\Delta: VJ^rY\ar_\delta VE \label{135}
\eeq
of the bundle morphism $\Delta$ (\ref{132}). Due to the canonical
isomorphism
\mar{134}\beq
VJ^rY= J^r(VY), \qquad \dot y^i_\La= (\dot y^i)_\La, \label{134}
\eeq
the bundle morphism (\ref{135}) is a $VE$-valued $r$-order
differential operator
\mar{136}\beq
V\Delta: J^r(VY)\ar_\Delta VE \label{136}
\eeq
on the vertical tangent bundle $VY$ of $V\to X$. It is called the vertical extension of the differential operator $\Delta$
(\ref{132}). Since $VE\to X$ is a vector bundle, the kernel of
this operator
\mar{137}\beq
V\gE=\Ker V\Delta  \label{137}
\eeq
defines an $r$-order differential equation on $VY$. With respect
to bundle coordinates $(x^\la, y^i, y^i_\La, \dot y^i, \dot
y^i_\La)$ on $J^r(VY)$, the differential equation (\ref{137}) is
locally given by equalities
\mar{140-2}\ben
&& \cE^A(x^\la, y^i_\La) =0, \label{140}\\
&& \dr_V\cE^A(x^\la, y^i_\La) =0, \label{141}\\
&& \dr_V=\dot y^i\dr_i + \dot y^i_\la\dr^\la_i +
+ \dot y^i_{\la\mu}\dr^{\la\mu}_i +\cdots , \label{142}
\een
where $d_V$ (\ref{142}) is the vertical derivative.

The equation $V\gE$ (\ref{137}) is called the deviation equation
(or the variation equation in the terminology of \cite{book09}).
Its part (\ref{140}) is the projection of this equation to $J^rY$,
and it is equivalent to the original equation (\ref{5.88}).
Therefore, a solution of the deviation equation (\ref{137}) is
given by a pair $(s,\psi)$ of a solution $s$ of the original
differential equation and a section $\psi$ of the pull-back bundle
$s^*VY\to X$ which obeys the linear differential equation
(\ref{141}). This section $\psi$ is called the Jacobi field.

In particular, if $Y\to X$ is an affine bundle modelled on a
vector bundle $\ol Y\to X$, there is the canonical isomorphism
\be
VY=Y\op\oplus_X \ol V,
\ee
and Jacobi fields $\psi$ are sections of a vector bundle $\ol Y\to
X$. For instance, if $Y$ is a vector bundle, then $\ol Y=Y$. In
the case of an affine bundle $Y$, it is readily observed that,
given a solution $(s,\psi)$ of the deviation equation $V\gE$
(\ref{137}), the sum $s+\psi$ obeys the original differential
equation $\gE$ with accuracy to terms linear in $\psi$. Therefore,
one can think of Jacobi fields $\psi$ as being deviations of
solutions of the original differential equation.

Let us note that any differential equation $\gE$ on a fibre bundle
can be written in the form (\ref{5.88}), and then the equalities
(\ref{140}) -- (\ref{141}) provide the corresponding deviation
equation $V\gE$.

Turn now to Lagrangian formalism on a fibre bundle $Y\to X$
\cite{book,book09,sard13}.

We use the fact that any exterior $m$-form $\phi$ on a fibre
bundle $Y\to X$ possesses a vertical extension
\mar{145}\beq
V\phi=\dr_V\phi \label{145}
\eeq
onto the vertical tangent bundle $VY$ of $Y\to X$. This is the
pull-back onto $VY\subset TY$ of its tangent extension $T\phi$
onto $TY$ defined by the equalities
\be
T\phi (\wt u_1,\ldots, \wt u_m)= u_{TY}\rfloor d(\phi (\wt
u_1,\ldots, \wt u_m))
\ee
for any vector fields $u_1,\ldots,u_m$ on $Y$, where $\wt u_a$ are
their functorial lift onto $TY$ and $u_{TY}$ is the Liouville
vector field on $TY$ \cite{book98}.

A $k$-order Lagrangian on a fibre bundle $Y$ is defined as a
density
\mar{144}\beq
L=\cL(x^\la, y^i, y^i_\La)d^nx, \qquad n=\di X, \label{144}
\eeq
on a $k$-order jet manifold $J^kY$. The kernel of the associated
Euler--Lagrange operator
\mar{150}\beq
\dl L=(\dr_i \cL + \op\sum_{\Lambda} (-1)^{|\Lambda|}d_\Lambda
\dr_i^\Lambda \cL)dy^i\w d^nx \label{150}
\eeq
are the $2k$-order Euler--Lagrange equations
\mar{151}\beq
\dr_i \cL + \op\sum_{\Lambda} (-1)^{|\Lambda|}d_\Lambda
\dr_i^\Lambda \cL=0 \label{151}
\eeq
on a fibre bundle $Y$.

Let us consider the vertical extension $VL$ (\ref{145}) of the
Lagrangian $L$ (\ref{144}) onto $VJ^kY=J^k(VY)$. It reads
\mar{146}\beq
VL=\dr_V\cL d^nx. \label{146}
\eeq
Therefore, one can think of $VL$ (\ref{146}) as being a $k$-order
Lagrangian on the vertical tangent bundle. It is easily verified
that the Euler--Lagrange operator $\dl VL$ of this Lagrangian $VL$
is the vertical extension $V\dl L$ (\ref{136}) of the
Euler--Lagrange operator $\dl L$ (\ref{150}) of a Lagrangian $L$.
Accordingly, the corresponding Euler--Lagrange equations are the
deviation of the Euler--Lagrange equations (\ref{151}).

Furthermore, the counterpart of a first order Lagrangian formalism
on a fibre bundle $Y\to X$ is polysymplectic Hamiltonian formalism
\cite{jpa99,book09} on the Legendre bundle
\mar{00}\beq
\Pi_Y =V^*Y\op\ot_Y(\op\w^nT^*X)\op\ot_YTX =
V^*Y\op\w_Y(\op\w^{n-1} T^*X) \label{00}
\eeq
provided with the holonomic coordinates $(x^\la, y^i, p^\la_i)$,
where the fibre coordinates $p^\la_i$ possess the transition
functions
\be
{p'}^\la_i = \det \left(\frac{\dr x^\ve}{\dr {x'}^\nu}\right)
\frac{\dr y^j}{\dr{y'}^i} \frac{\dr {x'}^\la}{\dr x^\m}p^\m_j.
\ee
The Legender bundle $\Pi_Y$ (\ref{00}) is provided with the
polysymplectic form
\be
\Om_Y =dp_i^\la\w dy^i\w \om\ot\dr_\la
\ee
and an exterior Hamiltonian form
\mar{b418}\beq
 H= p^\la_i dy^i\w \om_\la -\cH\om, \qquad \om=d^nx, \qquad\om_\la=\dr_\la\rfloor\om. \label{b418}
\eeq
This Hamiltonian form yields the covariant Hamilton equations
\mar{b4100}\beq
 y^i_\la=\dr^i_\la\cH, \qquad p^\la_{\la i}=-\dr_i\cH \label{b4100}
\eeq
on a fibre bundle $Y\to X$. A key point is that, due to the
canonical isomorphism $VV^*Y=V^*VY$, the vertical extension $VH$
(\ref{145}) of the Hamiltonian form $H$ (\ref{b418}) is a
Hamiltonian form
\be
VH=(\dot p^\la_i dy^i+ p^\la_i d\dot y^i)\w \om_\la -\dr_V\cH\om
\ee
on the Legendre bundle $\Pi_{VY}$ over the vertical tangent bundle
$VY$, and that the corresponding covariant Hamilton equations are
the deviation (\ref{137}) of the covariant Hamilton equations
(\ref{b4100}) \cite{book09}.

For instance, if $X=\mathbb R$, the above mentioned covariant
Hamiltonian formalism provides Hamiltonian formalism of
non-autonomous mechanics on a fibre bundle $Y\to \mathbb R$
\cite{book10,book98}. Its vertical extension has been considered
in application to mechanical systems with non-holonomic
constraints \cite{jmp99} and completely integrable systems
\cite{pla03,book10}. In particular, one can show that Jacobi
fields of a completely integrable Hamiltonian system of $m$
degrees of freedom make up an extended completely integrable
system of $2m$ degrees of freedom, where $m$ additional integrals
of motion characterize a relative motion \cite{pla03}.

\end{document}